# Multiferroic behavior confined by symmetry in EuTiO$_3$ films.

P. J. Ryan[1,9], G. E. Sterbinsky[1], Y. Choi[1], J. C. Woicik[2], Leyi Zhu[3], J. S. Jiang[3], J-H. Lee[4], D. G. Schlom[4,5], T. Birol[6], S. D. Brown[7], P. B. J. Thompson[7], P. S. Normile[8], J. Lang[1], and J.-W. Kim[1]

[1]*Advanced Photon Source, Argonne National Laboratory, Argonne, IL 60439, USA*
[2]*National Institute of Standards and Technology, Gaithersburg, MD 20899, USA*
[3]*Materials Science Division, Argonne National Laboratory, Argonne, IL 60439, USA*
[4]*Department of Materials Science and Engineering*
*Cornell University, Ithaca, NY 14853, USA*
[5]*Kavli Institute at Cornell for Nanoscale Science, Ithaca, NY 14853, USA*
[6]*Department of Chemical Engineering and Materials Science, 421 Washington Ave., SE, Minneapolis, MN 55455-0132, USA*
[7]*University of Liverpool, Dept. of Physics, Liverpool, L69 3BX, United Kingdom*
[8]*Instituto Regional de Investigación Científica Aplicada (IRICA) and Departamento de Física Aplicada, Universidad de Castilla-La Mancha, 13071 Ciudad Real, Spain*
[9]*School of Physical Sciences, Dublin City University, Dublin 9, Ireland.*

We have elucidated the spin, lattice, charge and orbital coupling mechanism underlying the multiferroic character in tensile strained EuTiO$_3$ films. Symmetry determined by oxygen octahedral tilting shapes the hybridization between the Eu 4f and Ti 3d orbitals and this inhibits predicted Ti displacement proper ferroelectricity. Instead, phonon softening emerges at low temperatures within the pseudo-cube (110) plane, orthogonal to the anticipated ferroelectric polarization symmetry. Additionally, the magnetic anisotropy is determined by orbital distortion through hybridization between the Ti 3d and typically isotropic Eu$^{2+}$ 4f. This unique scenario demonstrates the critical role symmetry plays in the coupling of order parameters defining multiferroic behaviour.

Uncovering the fundamental principles of strong magnetoelectric coupling, the intricate connections between multiple order parameters including spin, lattice, orbital, and charge often manifest as anomalous and intriguing physical phenomena (1,2). An interesting example demonstrating multiple parameter couplings is $EuTiO_3$ (ETO), which has been considered a parent candidate to explore multiferroic quantum criticality (3). In bulk, it is a quantum paraelectric-antiferromagnet (AFM) displaying strong magneto-dielectric behavior at the onset of its G-type AFM ordering (4,5). As illustrated in figure 1, ETO has co-existing ferromagnetic (FM) and AFM interactions described in the context of different pathways between Eu ions (6). Under tensile strain, ETO thin films have demonstrated FM order, which was predicted and shown to be coupled to ferroelectric order induced by Ti non-central symmetry (7,8).

Competing magnetic interactions and a large and diverging dielectric constant at low temperatures indicating proximity to a ferroelectric transition make ETO a strong parent candidate for multiferroic quantum critical behavior. It is theoretically predicted that a combination of chemical pressure, magnetic dilution and strain may engineer both magnetic and ferroelectric quantum criticality to coincide (9). However, in order to explore such exotic phenomena understanding the coupling between the order parameters of the parent compound is crucial. In particular, the role symmetry plays in shaping the orbital states of both Eu and Ti and their interactions will ultimately determine how both quantum critical points behave and interact.

In this work, we present a study of the relationship between the electric and magnetic polarizations with structural symmetry as determined by the octahedral ($O_h$) rotation pattern in tensile strained ETO epitaxial thin films. We explain how the FM order exhibits uniaxial anisotropy and remarkably why the system resists breaking central symmetry. The oxygen $O_h$ tilt pattern alters the orbital topology consisting of Eu (4f) - O (2p) - Ti (3d) bands. In doing so, the tilt differentiates between the $(110)_{pc}$ and $(-110)_{pc}$ planes generating an azimuthally dependent inter-orbital distortion of the Eu 4f states. In the $d^0$ titanite perovskites, the oxygen $O_h$ tilt would generally lead to central symmetry-breaking along the $O_h$ rotational vector (10,11). Instead a significant softening of the Slater mode is observed within the plane orthogonal to the anticipated FE polarization direction (10). Polarized x-ray absorption near edge spectroscopy at the Ti K edge shows the correlated Ti phonon softening at low temperatures in this plane where crystal symmetry forbids central symmetry breaking. In addition employing resonant x-ray magnetic scattering, we relate both spin and phonon anisotropy with structural symmetry. In aggregate, these measurements reveal the relative oriented coupling behavior of the symmetry with spin, lattice, and orbital parameters of the film.

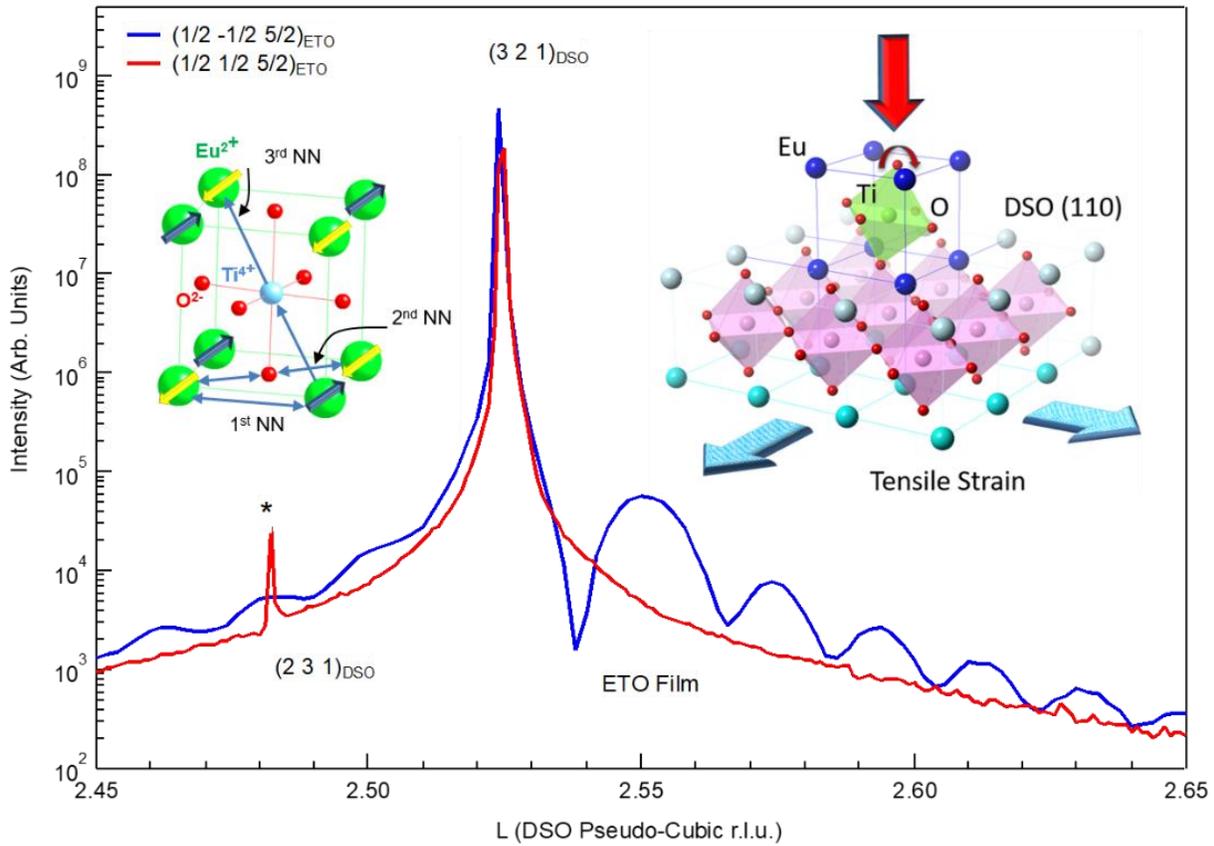

Figure 1. Reciprocal unit L scans of half order reflections resulting from the oxygen octahedral tilt pattern. Absence of the half order (1/2 -1/2 5/2) indicates that the film has a single type tilt domain. Cartoons illustrating (left) ETO pseudo cubic unit cell showing G-AFM order with all three co-existing spin ordering interactions and (right) the biaxial strain and single oxygen related tilt pattern imposed by the DSO (110) substrate.

The ETO epitaxial 25nm film, grown by reactive molecular-beam epitaxy, is fully constrained to the $DyScO_3$ (DSO)(110)$_o$ (Orthorhombic) substrate (8). The orthorhombic symmetry of the DSO substrate induces a single $O_h$ rotation pattern in the ETO film as identified by x-ray diffraction measurements of half order reflections illustrated in figure 1 (12). As we discussed previously (11), the tensile strain leads to the metastable state, Imma, $(a^-a^-c^0)$ (12) with symmetry breaking between [110] and [-110] through $O_h$ tilting. There are two possible tilt domains, along [110] and [-110] directions. The half order (½ ½ L/2) reflections are allowed for the [110] $O_h$ tilt domain but not for the [-110] domain. The absence of (½ ½ 5/2) reflection therefore verifies the single tilt pattern.

The Ti atom is of particular importance to the electronic band characteristics of this material and considered responsible for ferroelectricity in the titanate-perovskite systems (15). Additionally, it is a key element underlying the multiferroic, magneto-electric and possible multiferroic-quantum critical characteristics of $EuTiO_3$ (3,5-10), as it mediates the G-type AFM super-exchange mechanism between the Eu (4f) states via the Ti 3d orbitals (16). In fact the localized 4f spins hybridize with the Eu (5d), Ti (3d) and O (2p) states, illustrating the potential of strong inter-atomic coupling phenomena (16). To investigate the Ti behavior, we employed linear-polarized x-ray absorption spectroscopy at the Ti K edge (Figure 2) to study the configuration of Ti electronic states. The K edge is dominated by the allowed dipole transition from 1s to 4p orbitals while the 1s to 3d transition is forbidden. However *p*-character is introduced to the 3d states through orbital hybridization that occurs with static or dynamic symmetry breaking, which gives rise to pre-edge features. For this reason the pre-edge $e_g$ intensity is proportional to the square of the component of the Ti displacement along the incoming beam polarization (17- 19).

In figure 2b, the inset cartoon illustrates the grazing incidence experimental geometry. The lack of an anomalously large change in the intensity of the pre-edge $e_g$ peak with temperature suggests that there is not static central symmetry breaking. This then rules out the picture of simple proper FE behavior via a B-site shift. However, in figure 2 there are subtle but significant changes to the $e_g$ intensity. These effects relate to modulus changes of the phonon modes (20). Figure 2b presents a comparison of the $e_g$ intensity between incoming photon polarizations, e along the film in-plane [1 0 0] and out-of-plane [0 0 1] directions, at room temperature and at 15K. At 15K the reversal of the in/out-of-plane $e_g$ intensity demonstrates an unexpected increase of the modulus of the position normal to the film as illustrated in the cartoon model, fig. 2c while the in-plane phonon modulus decreases with temperature.

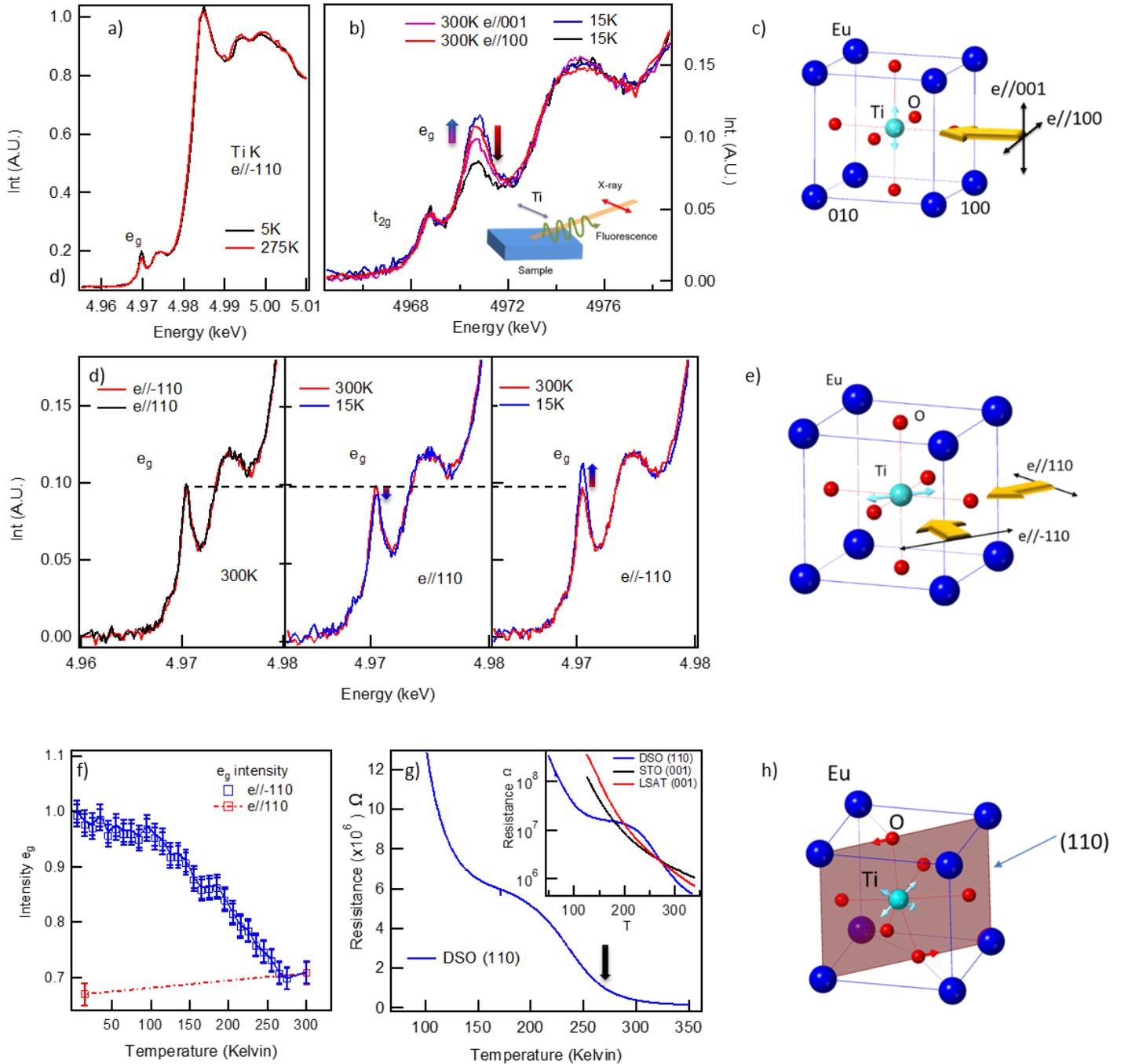

Figure 2. a) Ti K-edge near edge spectra with linear polarization at 5 and 275K. b) Expanded plots of the spectral intensities of the $e_g$ pre-edge peak contrasting the temperature dependence of polarization parallel to both in-plane [100] and out-of-plane [001] as depicted in the inset. c) Cartoon illustration of the relative incident polarization (yellow arrows) with low temperature phonon softening effect along [001] (cyan arrow). d) Azimuthal-temperature dependence of $e_g$ pre-edge intensity along both e//[110] and e//[-110]. e) Cartoon displaying the relative incident geometries with arrow indicating increased softening. f) Temperature dependence of the $e_g$ pre-edge intensity with the polarization along the [110] and [-110]. g) Temperature dependence of the in-plane film resistance. Inset shows the comparison between other strain states. h) Cartoon showing the aggregate relative orientation of Ti phonon softening along the cube diagonal with respect to the oxygen tilting within the (110) plane.

Similar in-plane azimuthal comparisons are shown in figure 2d. The pre-edge $e_g$ intensity change derives primarily from a temperature effect and the intensities are indistinguishable between e//[110] and e//[-110]. However, at 15K the intensity of e//[-110] increases and is attributed to a phonon softening along this direction (Fig. 2e), whereas along the [110] the anticipated decrease is observed. Having an increased $e_g$ intensity both along [001] and [-110] suggests a cube diagonal [-111] (or [1-11]) net softening phenomenon. The temperature dependence is very similar to the published SHG results (8). An uptick in resistance is measured at roughly the same temperature (Fig. 2g) as the $e_g$ intensity in figure 2f. Recent DFT calculations propose net band gap widening with increasing O 2p and Ti 3d orbital mixing in biaxially strained $SrTiO_3$ films (17). Lowering the symmetry through ferroic distortions by either oxygen $O_h$ rotation or Ti displacement, the conduction (O 2p) and valence states (Ti 3d) are allowed to mix as the distinction between bond phase parity is removed resulting in charge repulsion (22). Although dynamic, the Ti displacement along the cube diagonal widens the gap resulting in a clear uptick in resistance at ~270K. This behavior demonstrates that the band edge active orbitals Ti (3d) and O (2p) are tightly coupled and respond to the amplitude of atomic vibrations.

The lack of central symmetry breaking can be understood as the effect of cross-gap hybridization between the empty transition-d (cation) band and the occupied O 2p band (20). Any Ti shift towards the oxygen increases the Ti-O hybridization further increasing the gap between the unoccupied states (higher) and the occupied states (lower) (23). Normally, with partial occupation of the d-state there is an additional energy cost pushing the unoccupied d-states higher. This reduces the probability of FE order (24). This model also explains the magneto-dielectric spin-lattice coupling in ETO (6). Consequently, a small residual occupation of the Ti (3d) orbital due to hybridization inhibits symmetry breaking. Absent the half-filled Eu (4f) orbitals, the Ti would likely displace and result in proper FE order (6).

The seminal work of Zhong and Vanderbilt (25) predicted the emergence of Ti FE order in $SrTiO_3$ under negative iso-pressure (increased volume) where the Ti displacement displays both in-plane (x,y) and out-of-plane (z) components. Unlike the isotropic negative pressure considered in these calculations, biaxial tensile strain generates an $O_h$ rotation resulting in a symmetry breaking effect ($a^-a^-c^o$). Covalent bonding between the A-site (Eu) and oxygen strengthens the $O_h$ rotational order (26) and subsequently increases the orbital interaction and hybridization between the Eu (4f) and Ti (3d) states. (27) Furthermore, the single $O_h$ tilt pattern in this film differentiates between the inter-orbital hybridization within the (110) and (-110) planes. Strictly speaking, the symmetry of the film is $a^-b^-c^0$ rather than $a^-a^-c^0$ because the DSO substrate does not have equal in-plane lattices. Our measurements support that the cube diagonal within the (110) plane is the phonon softening direction, which is not the expected FE polarization direction in this crystallographic group. Symmetrically the Ti displacement in the tensile environment is anticipated to be along the $O_h$ rotational axis, parallel to the (110), i.e. $a_+^-b_+^-c_0^o$ (8Cm). However, the measured softening occurs along [-111] indicative of $a_-^-b_+^-c_+^o$, along which a static displacement is not symmetrically allowed. (14)

The symmetry of the hybridization landscape involving all three elements is evidently broken with the $O_h$ unidirectional tilting, affecting the degree of hybridization between the Eu (4f) and Ti (3d) within both (110) and (-110) planes. The instability of the Ti phonon is dependent on the effective charge sharing (hybridization), the more net charge resident on the Ti orbitals the stiffer the phonon character. The tilting of the O towards the Eu atom repels the charges within this plane (110) reducing the effective hybridization and allowing a degree of softening. However full freezing is, ultimately inhibited by symmetry.

Simple Ti displacement is unlikely to be the origin of the reported FE behavior (7), however, the A-site may alternatively be the origin of the polar effect as has been reported driving the improper FE order in $NdNiO_3$ films (28). In $EuTiO_3$, the smaller size of the $Eu^{2+}$ ion compared to typical B-site ferroelectrics such as $BaTiO_3$ makes it likely that the Eu is more active in any potential ferroelectricity than Ti. In fact, analysis of neutron diffraction estimating atomic displacement factors suggests a Eu *delocalization* in bulk single crystal $EuTiO_3$ (29). Interestingly, as a consequence of twinned $O_h$ antiferrodistortive phases A-site driven anti-ferroelectric order was calculated in similarly tensile strained $EuTiO_3$ (30), and potentially demonstrated in bulk ETO (31).

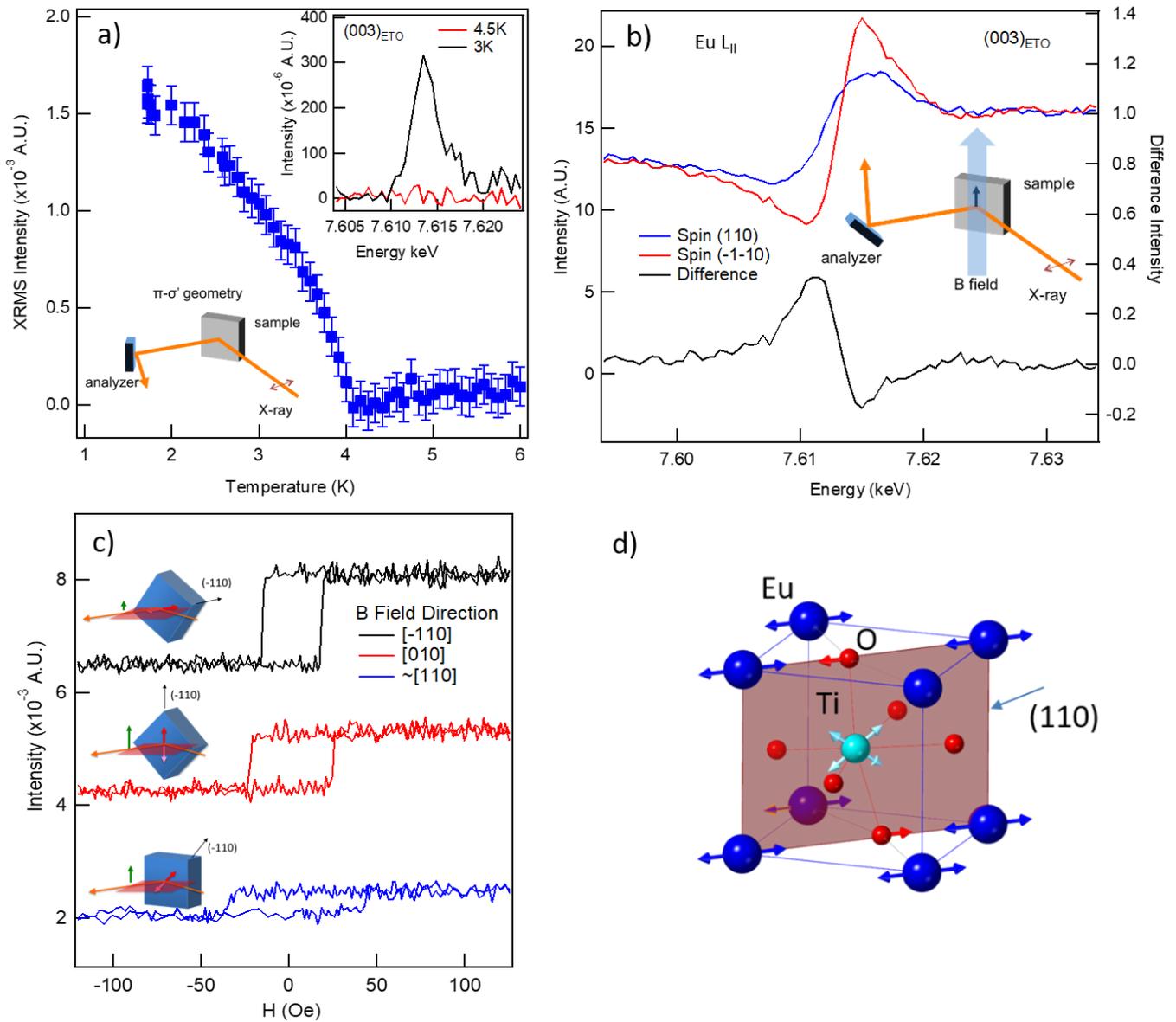

*Figure 3. a) Temperature dependent XRMS scattering intensity of the $(003)_{ETO}$ reflection at the Eu $L_{II}$ edge indicative of FM spin order. The inset of bottom left corner illustrates the horizontal scattering experimental configuration (π-σ) and the top right shows resonant enhancement (background subtracted) above and below $T_c$. b) Energy scans at the $(003)_{ETO}$ reflection in (π-π) geometry where the charge scattering intensity can interfere with the magnetic scattering. The in and out of phase interferences of magnetic and charge scattering are dependent on the aligned spin direction. c) Field dependent sweeps with fixed energy in (π-π) showing clear FM hysteresis loops. Data is offset for clarity. With the applied field along the easy magnetic axis [-110] the hysteresis loops show minimum coercivity. Along the [010] and close to the [110] the intensity difference between positive and negative field directions decreases and the loop broadens. Inset images illustrate regardless of changes of the field direction by sample rotation the spins stay within the easy axis [-110]. d) A cartoon model illustrating the relative orientation of the oxygen $O_h$ tilting, Ti phonon softening and uniaxial magnetic anisotropy.*

The tensile strain induced distortion drives the film into a FM state as previously demonstrated (8). In addition to Ti displacement (7), DFT describes increasing volume alone driving this spin transition (32). Interestingly, the calculated volume for isotropic expansion ~248 (Å$^3$) agrees well with the measured volume of the tensile strained state here of 243.32 (Å$^3$). The Eu$^{2+}$ has a spherical 4f$^7$ spin configuration, which has nominally zero orbital angular momentum. This leads to an immunity from crystal field splitting effects that cause single ion magnetic anisotropy. Instead, multi-directional hybridization of f-d-p (6,16) orbitals driven by oxygen octahedral tilts, can distort the local Eu 4f electronic configuration, resulting in magnetic anisotropy. Here, we conclusively link the uniaxial magnetic anisotropy with its easy axis along the [-110], with the single symmetry orientation and lattice behavior by employing x-ray magnetic scattering (XRMS) to concomitantly probe the corresponding Eu spin orientation.

The appearance of an integer reflection (003) at the Eu L$_{II}$ edge in figure 3 confirms the FM spin order (33). In figure 3a, the temperature dependence of magnetic intensity illustrates a T$_c$ of ~4.2K and a critical exponent β of ~0.36. Employing (π-π) polarization analysis at the (003) reflection (inset Fig 3b), the intensity of both charge and magnetic scattering is comparable, allowing phase interference. Consequently the measurement is sensitive to the spin direction with respect to the lattice (34). The difference in energy dependence of the (003) reflection with the applied magnetic field along the [-110] and [1-10] directions (Fig. 3b) illustrates the phase relationship between the magnetic and charge scattering. Figure 3c displays field loops at fixed energy 7.612keV at 3.2K (above the DSO T$_N$) applying the field along the ±[-110], ± [010], and ~ ±[110] directions. The spins remain within [-110] direction regardless of field direction as illustrated in fig 3c inset where the effective magnetic scattering cross section is represented by the green arrow. With Q vector rotation the magnetic cross section decreases as the spin easy-axis approaches the scattering plane. The loops broaden as the effective field that requires flipping the spins increases. In figure 3d inset, the conclusion of all three measurements are presented in schematic form, the oxygen O$_h$ single tilt domain pattern, with uniaxial FM spin anisotropy below T$_c$ 4.2K and finally the increased in-plane Ti fluctuations below ~270K, all co-align within the (110) plane.

In summary, we have established how a unique symmetry driven by oxygen octahedral tilt shapes the hybridization between the Eu 4f and Ti 3d orbitals in ETO films. This substrate-imprinted symmetry inhibits predicted Ti displacement proper ferroelectricity, instead, phonon softening critical to ferroelectric order emerges at low temperatures. Confined within the (110) plane there exists significant orbital distortion instigated through the O$_h$ oxygen displacement shaping both the Ti phonon landscape and the Eu 4f orbitals. Hybridization between Ti 3d and Eu 4f states facilitates the effective charge transfer (sharing) between the two, leading to expected ferroelectric order suppression. The single oxygen octahedral tilt domain fortuitously creates an ideal opportunity to study the underlying nature of how spin, lattice, orbital and charge order parameters couple. In doing so, we show how 4f spin structure is set by inter-orbital shaping as determined by intricately entangled order parameters confined ultimately by symmetry. Our findings will provide guidance in varying external tuning parameters to study multiferroic quantum criticality in ETO related systems.

*Acknowledgements*
*Work at Argonne and use of beamline 6-ID-B at the Advanced Photon Source at Argonne was supported by the U. S. Department of Energy, Office of Science, Office of Basic Energy Sciences under Contract No. DE-AC02-06CH11357. The EPSRC-funded XMaS beamline at the ESRF is directed by M.J. Cooper, C.A. Lucas and T.P.A. Hase. We are grateful to O. Bikondoa, D. Wermeille, and L. Bouchenoire for their invaluable assistance and to S. Beaufoy and J. Kervin for additional XMaS support. The work at Cornell was supported by the US Department of Energy, Office of Basic Energy Sciences, Division of Materials Sciences and Engineering, under Award No. DE-SC0002334. Substrate preparation was performed in part at the Cornell NanoScale Facility, a member of the National Nanotechnology Coordinated Infrastructure (NNCI), which is supported by the National Science Foundation (Grant ECCS-1542081). Use of beamline X23-A2 at the National Synchrotron Light Source was supported by the U.S. Department of Energy, Office of Basic Energy Sciences, under Contract No. DE-AC02-98CH10886. Additional support was provided by the National Institute of Standards and Technology*